\documentclass[twocolumn,showpacs,preprintnumbers,amsmath,amssymb]{revtex4}
\usepackage{graphicx}
\usepackage{epsfig}
\usepackage{dcolumn}
\usepackage{bm}
\newcommand{\BR}{\mathcal{B}}
\newcommand{\psipto}{\psi(2S)\rightarrow \pi^+\pi^- J/\psi}
\newcommand{\pnb}{p\bar{n}\pi^-}

\newcommand{\pppr}{\pi^+\pi^-p\bar{p}}

\newcommand{\LL}{\ell^+\ell^-}
\newcommand{\EE}{e^+e^-}
\newcommand{\MM}{\mu^+\mu^-}

\newcommand{\pp}{\pi^+\pi^-}

\newcommand{\ppb}{p\bar{p}}

\newcommand{\ppuu}{\pi^+\pi^-\mu^+\mu^-}

\newcommand{\nnb}{n\bar{n}}

\newcommand{\psp}{\psi(2S)}
\newcommand{\jpsi}{J/\psi}
\newcommand{\ar}{\rightarrow}

\newcommand{\bfg}{\begin{figure}}
\newcommand{\efg}{\end{figure}}
\newcommand{\bitm}{\begin{itemize}}
\newcommand{\eitm}{\end{itemize}}
\newcommand{\bnum}{\begin{enumerate}}
\newcommand{\enum}{\end{enumerate}}
\newcommand{\btbl}{\begin{table}}
\newcommand{\etbl}{\end{table}}
\newcommand{\btbu}{\begin{tabular}}
\newcommand{\etbu}{\end{tabular}}
\newcommand{\bcl}{\begin{center}}
\newcommand{\ecl}{\end{center}}
\newcommand{\bbt}{\bibitem}
\newcommand{\beq}{\begin{equation}}
\newcommand{\eeq}{\end{equation}}
\newcommand{\beqr}{\begin{eqnarray}}
\newcommand{\eeqr}{\end{eqnarray}}

\begin{document}
\title{\boldmath Search for Invisible Decays of the $\jpsi$ in $\psipto$}
\author{
M.~Ablikim$^{1}$,              J.~Z.~Bai$^{1}$,               Y.~Ban$^{12}$,
X.~Cai$^{1}$,                  H.~F.~Chen$^{17}$,
H.~S.~Chen$^{1}$,              H.~X.~Chen$^{1}$,              J.~C.~Chen$^{1}$,
Jin~Chen$^{1}$,                Y.~B.~Chen$^{1}$
% ...
\\
\vspace{0.2cm}
(BES Collaboration)\\
\vspace{0.2cm}
{\it
$^{1}$ Institute of High Energy Physics, Beijing 100049, People's Republic of China\\
% ...
$^{12}$ Peking University, Beijing 100871, People's Republic of China\\
% ...
$^{17}$ University of Science and Technology of China, Hefei 230026, People's Republic of China\\
% ...
}}

\date{\today}

\begin{abstract}
  Using  $\psipto$ events in a sample of  
  14.0 million $\psp$ decays collected with the BES-II detector, a
  search for the decay of the $\jpsi$ to invisible final states
  is performed.  The $\jpsi$ peak in the distribution of
  masses recoiling against the $\pp$ is used to tag $\jpsi$ invisible
  decays. No signal is found, and an upper limit at the 90\% confidence
  level is determined to be $1.2\times 10^{-2}$ for the ratio
  $\frac{{\cal B}(\jpsi\ar \mbox{invisible})}{{\cal B}(\jpsi\ar\MM)}$. 
  This is the first search for $\jpsi$ decays to invisible final
  states.
\end{abstract}
\maketitle
Invisible decays of quarkonium states such as $\jpsi$ and
$\Upsilon$, etc., offer a window into what may lie beyond the
standard model (SM)~\cite{0,00,000}. This is because, aside
from neutrinos, the SM includes no other invisible particles that
these states can decay into. In the SM, the predicted branching
fraction for $\jpsi \rightarrow \nu \overline{\nu}$ is
\begin{equation}
\BR(\jpsi \rightarrow \nu \overline{\nu}) = 4.54 \times 10^{-7}
\times \BR(\jpsi \rightarrow e^+e^-)
 \label{eq:sm_predict}
\end{equation}
\noindent
with a small uncertainty (2-3\%)~\cite{0,1}.  However, new physics beyond
the SM might enhance the branching fraction of $\jpsi$ invisible
decays. One possibility is the decay into light dark matter (LDM)
particles mediated by a new, electrically neutral spin-1 gauge boson $U$, 
which
could significantly increase the invisible decay rate~\cite{2}. On the
other hand, $C$-even operators $\bar{q}q$ or $\bar{q} \gamma_5 q$ do not
contribute to invisible decays of the J/$\psi$~\cite{00,000}. Thus,
there would be no 
contributions from a possible scalar or pseudoscalar exchange
to the decay of J/$\psi$~\cite{000,3}.

Astronomical observations of a bright 511~keV $\gamma$-ray
line from the galactic bulge have been reported by the SPI
spectrometer on the International Gamma-Ray Astrophysics Lab
(INTEGRAL) satellite~\cite{4}. The corresponding galactic positron
flux, as well as the smooth symmetric morphology of the 511~keV
emission, may be interpreted as originating from the annihilation
of LDM particles into $e^+e^-$ pairs~\cite{2}. It is
of interest to search for such light invisible particles in collider
experiments.  CLEO reported an upper bound on $\Upsilon(1S) \rightarrow
\gamma + \mbox{invisible}$, which is sensitive to dark matter
candidates lighter than about 3 GeV/$c^2$ and also provides an upper
limit on the axial coupling of a new $U$ boson to the $b$
quark~\cite{5}. Recently, both CLEO and Belle reported upper limits on
$\Upsilon(1S) \rightarrow \mbox{invisible}$ decays~\cite{6}. The first
experimental limits on invisible decays of the $\eta$ and
$\eta^\prime$ mesons have  been reported by the BES
Collaboration~\cite{7}; these can be used to constrain the mass of the
LDM particles and couplings of new bosons to light quarks. Here, we
present the first search for $\jpsi$ decays to invisible final states.

BES-II is the upgraded version of the BES-I detector~\cite{8}. A
12-layer vertex chamber (VC) surrounding the beam pipe provides
trigger and position information. This detector efficiently detects
the presence of charged tracks over 97\% of the total solid angle.
A forty-layer main drift chamber
(MDC), located radially outside the VC, provides trajectory and energy
loss ($dE/dx$) information for charged tracks over $85\%$ of the total
solid angle.  The momentum resolution is $\sigma _p/p = 0.017
\sqrt{1+p^2}$ ($p$ in $\hbox{\rm GeV}/c$), and the $dE/dx$ resolution
for hadron tracks is $\sim 8\%$.  An array of 48 scintillation
counters surrounding the MDC measures the time-of-flight (TOF) of
charged tracks with a resolution of $\sim 200$ ps for hadrons. Outside
of the TOF counters is a 12-radiation-length barrel shower counter
(BSC) composed of gas tubes interleaved with lead sheets. This
measures the energies of electrons and photons over $\sim 80\%$ of the
total solid angle with an energy resolution of
$\sigma_E/E=22\%/\sqrt{E}$ ($E$ in GeV).  Outside of the solenoidal
coil, which provides a 0.4~Tesla magnetic field over the tracking
volume, is an iron flux return that is instrumented with three double
layers of counters that identify muons of momenta greater than 0.5
GeV/$c$.

In order to detect invisible $\jpsi$ decays, we use $\psipto$ and
infer the presence of the $\jpsi$ from the $\jpsi$ peak in the
distribution of mass recoiling against the $\pi^+\pi^-$. In this
analysis, we use a 19.72 pb$^{-1}$ data sample collected at the peak of 
the $\psp$ resonance and a 6.42 pb$^{-1}$ data sample collected 
off-resonance at a center-of-mass energy of $\sqrt{s} = 3.65$ GeV.
The data were recorded in the BES-II
detector operating at the Beijing Electron Positron Collider (BEPC).
The data sample contains 14.0 million $\psp$ decays.

In the search for $\jpsi$ invisible decays, invisible means nothing
besides the two pions is seen in either the tracking or calorimetry
systems of the
detector. The charged-track trigger in BES-II requires at least one hit in 
the 48 barrel TOF counter array, one track in the VC and MDC, and at least
100 MeV of energy deposit in the BSC~\cite{8, 9}.  This trigger is
sensitive to two soft pions from the $\psi(2S)$ decay of the signal
events.

Events with only two charged tracks with zero net charge are selected.  
Each charged track is
required to be well fitted by a helix and to have a polar angle,
$\theta$, within the fiducial region $|\cos\theta|<0.8$. To ensure
that the tracks originate from the interaction region, we require
$V_{xy}=\sqrt{V_x^2+V_y^2}<2$ cm and $|V_z|<20$ cm, where $V_x$,
$V_y$, and $V_z$ are the $x, y$, and $z$ coordinates of the point of
closest approach of each charged track to the beam axis. In addition,
the momentum of each charged track must be less than 0.4
GeV/$c$. Particle identification (PID) is performed with the combined
TOF and $dE/dx$ information, and both charged tracks must be identified
as pions.  The invariant mass of the $\pp$ pair is required to be
larger than 0.35 GeV/$c^2$.

In BES-II, high momentum charged tracks with $|\cos\theta|<0.8$ 
traverse the full radial extent of the MDC are detected and reconstructed 
efficiently.  Tracks with lower momentum and/or at larger angles miss some 
or all of the MDC layers and may not be reliably reconstructed. However, 
tracks that originate from the center of the interaction point with 
a polar angle in the range $|\cos\theta|<0.97$  
penetrate at least six layers of the inner vertex chamber;  Although these 
tracks are not fully reconstructed, their presence can be inferred with high
efficiency from track segments reconstructed from three or more VC hits.  
Events with indications of additional charged tracks 
anywhere in the region of angular coverage of the vertex chamber are rejected.

Each reconstructed BSC cluster is required to have an energy greater than 10 
MeV and to have a cluster profile consistent with that of a shower in the 
BSC. The number of unassociated neutral clusters, which do not match with 
either charged track in the MDC, is required to be zero in order to suppress 
backgrounds from $\jpsi$ decaying into neutral final states.

Figure~\ref{fit} shows the distribution of masses recoiling against
the $\pp$ pair for candidate events.
A clear $\jpsi$ peak is evident for the data taken at the $\psi(2S)$,
while the smooth background under the $\jpsi$ peak from the
QED contribution is consistent with the distribution obtained with
off-resonance data at $\sqrt{s}=3.65$ GeV after normalizing to
the luminosity at $\sqrt s=M_{\psp}$.

\bfg 
\centerline{\psfig{file=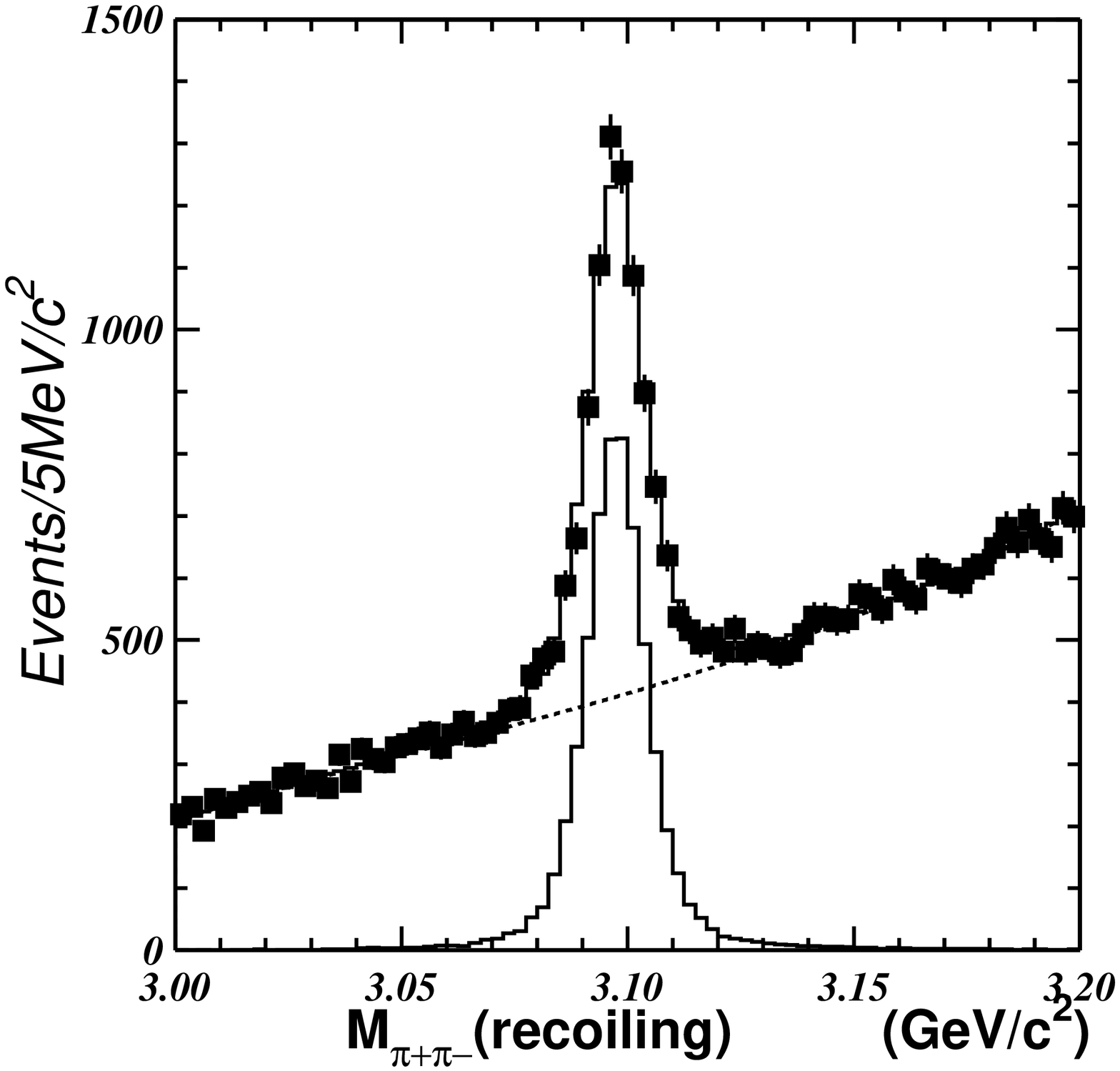,width=6cm,height=6cm}}
\caption{Distribution of masses recoiling against the $\pp$ for
  $\psi(2S) \to \pi^+ \pi^- J/\psi, J/\psi \to $
  invisible candidate events. Dots with error bars denote data.  The
  lower histogram is the signal shape from $\psi(2S) \to \pi^+ \pi^-
  J/\psi, J/\psi \to \mu^+ \mu^-$ events.  The
  upper histogram and the dashed curve are from the fit.}
\label{fit}
\efg

By measuring the ratio of invisible $\jpsi$ decays to 
$\jpsi\ar\MM$,
many uncertainties cancel, including those related to the
number of $\jpsi$ decays, the soft pion tracking efficiencies,
and the zero neutral cluster requirement.
First, two soft pions are chosen with
the same selection criteria as used for $\psipto,\jpsi\ar
\mbox{invisible}$. Then, we require that there are two muons in
addition to the two soft pions.  Two selection criteria are used to
identify muons. One is that the momentum of each track is larger than 
0.7~GeV/$c$. The other is that R, which is defined as
$$R=\sqrt{\left(\frac{E_{sc1}}{p_1}-1\right)^2
+\left(\frac{E_{sc2}}{p_2}-1\right)^2},$$ be larger than 1.0 in order
to reject $\jpsi\ar\EE$ events.  Here, $E_{sc1}$ and $E_{sc2}$ denote
the deposited energies in the BSC, and $p_1$ and $p_2$ denote their
momenta.  Figure~\ref{R} (a) shows the $R$ distributions for data and
Monte Carlo (MC) simulated events.  For each event, the total energy
of the four charged tracks, $E_{tot}$, is required to be larger than
3.6 GeV, as shown in Fig.~\ref{R} (b).  The small difference between
the $E_{tot}$ distributions for data and MC simulation is caused by
imperfect simulation of soft pions, an effect that is considered in
the systematic error study.  The requirement on the number of
unassociated neutral clusters or extra charged tracks is the same as that 
for $\psipto,\jpsi\ar \mbox{invisible}$ so that the systematic
uncertainties cancel in the ratio. Figure~\ref{ppuu} shows the
$M_{\pp}^{\mbox{recoil}}$ distribution for $\ppuu$ final states for
data and MC simulation after the above selection. Compared to the data
distribution, the MC sample has a shift of 1.0 MeV/c$^2$, which is
also caused by the imperfect simulation of soft pions. The number of
$\psi(2S) \to \pi^+ \pi^- J/\psi, J/\psi \to \mu^+ \mu^-$ events
selected is $43429\pm 208$ where the error is statistical, and the
corresponding MC-determined detection efficiency is 14.5\%.

\bfg[htpb]
\centerline{\psfig{file=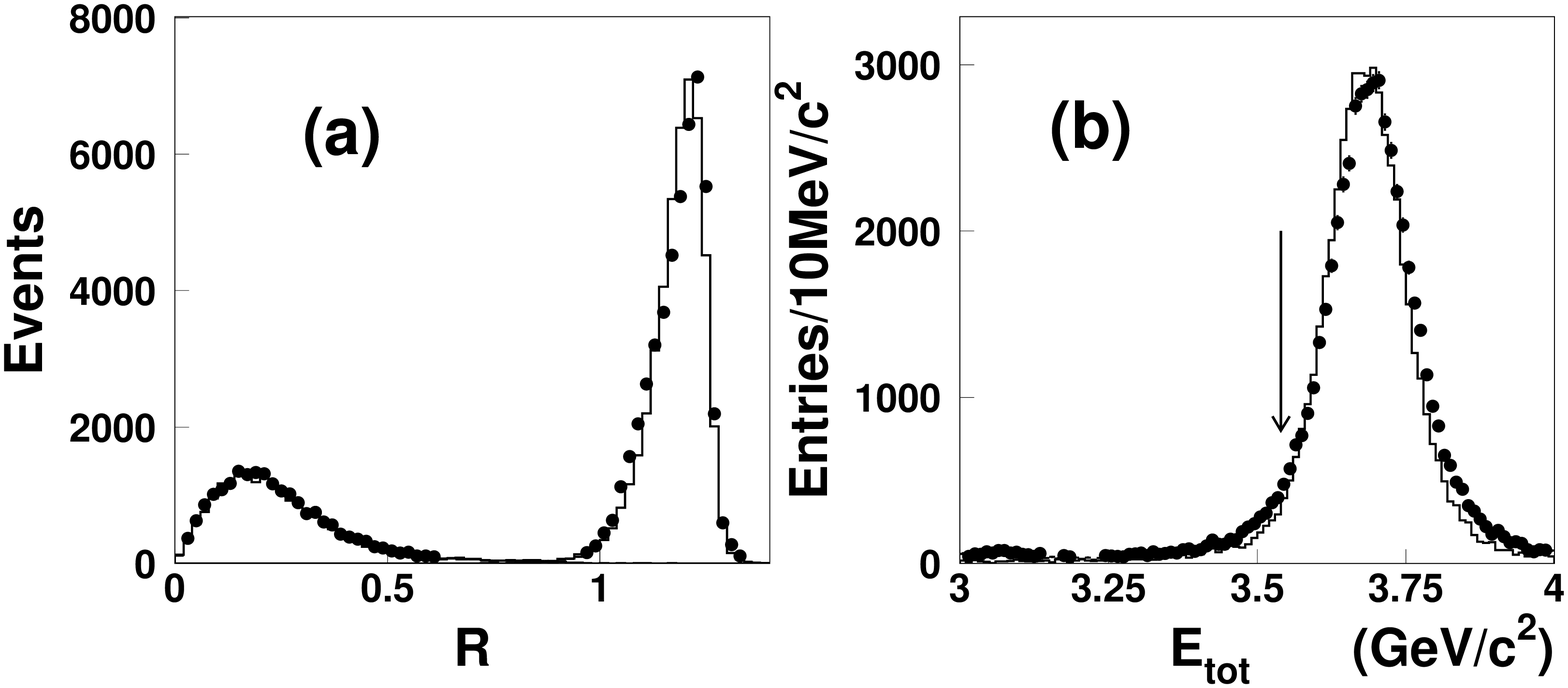,width=8cm,height=4.5cm}}
\caption{(a) Distributions of $R$ for the two non-pion tracks in
$\psipto,\jpsi\ar\LL$. (b) Distributions of $E_{tot}$ for
$\psipto,\jpsi\ar\MM$. The histogram denotes
MC and dots with error bars denote data.}
\label{R}
\efg

\bfg \centerline{\psfig{file=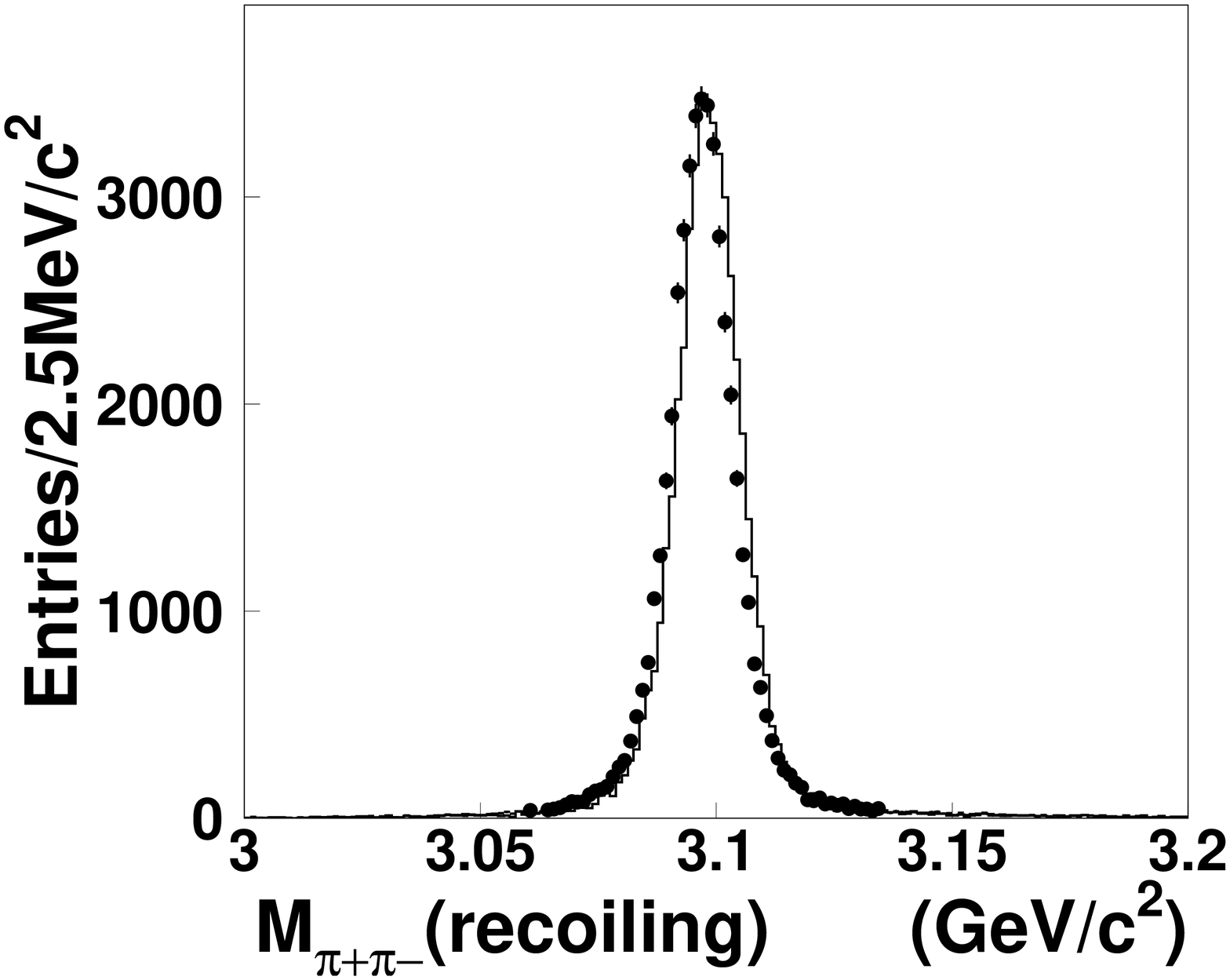,width=6cm,height=5cm}}
\caption{Distribution of mass recoiling against the $\pp$ pair for
  $\psp \rightarrow \ppuu$ candidate events. The histogram denotes
MC simulation and dots with errors denote data. }
\label{ppuu}
\efg

Exclusive channels that are potential backgrounds are studied using
full MC simulations in order to determine their contamination. The
sources of the backgrounds can be divided into two classes. Class I is
mainly from final states without a $J/\psi$, such as the continuum
processes $e^+ e^- \rightarrow q \bar{q}$ and $\tau^+\tau^-$ ($q= u$,
$d$ and $s$ quarks). The smooth backgrounds under the $J/\psi$ peak in
Fig.~\ref{fit} are from "Class I".  Class II is from $\psipto$, where
the $J/\psi$ decays into other modes than the true invisible final
state. For this case, the $J/\psi$ decay products are either outside of
the detector acceptance, or are inside but are undetected. For these
kinds of backgrounds, there is a $\jpsi$ peak in the
$M^{\mbox{recoil}}_{\pi^+\pi^-}$ distribution that is identical to the
invisible signal. The largest sources of ``peaking backgrounds"
are from the J/$\psi$ decays to $\LL(\ell=e,\mu)$, where
the two tracks tend to be back to back, so when one track escapes
into the forward insensitive region, the other track tends to escape
into the backward insensitive region. This is a geometric effect and its 
size is determined from Monte Carlo simulation. Other sources of backgrounds 
are from decays to $\nnb$, where the neutron and anti-neutron leave no hit
information in the detector.  An additional potential background comes
from $\jpsi$ decays to $\nnb\pi^0$, in which neither of the two photons
from $\pi^0$ decay are within the detector acceptance and the
neutron and antineutron either have no hit information in the detector
or miss the sensitive region of the detector.

The systematic uncertainties on the number of expected background
events caused by the selection criteria are estimated using special
event samples. For the dominant backgrounds from $\jpsi$ decays to
$\LL$, the uncertainty is mainly from the difference between data
and MC simulation of the contamination from two missing leptons. 
We determine this systematic
error using $\jpsi\ar K_S K^{\pm}\pi^{\mp}$, in which only the $K_S$
and the charged pion are required in event selection.   This provides
a large sample of ($\simeq$38.6K) tagged charged kaons in the detector.
With these, we measure the probability for a charged kaon to be 
``invisible,'' {\it i.e.,} to evade the extra track requirements, 
to be $(3.49\pm 0.09)\%$.  A MC simulation of this process gives an 
invisibility probability of $(3.13\pm 0.07)\%$. The difference of 
$\simeq 10\%$ between the measured and simulated probabilities
is taken as a systematic error for peaking backgrounds from $\jpsi$ 
decays to $\LL$ and $\ppb$.

In estimating the expected background contamination from $\jpsi\ar\nnb
$, we assume the branching fraction and its corresponding uncertainty
for $\jpsi\ar\nnb$ are the same as those for $\jpsi\ar\ppb$ as is expected 
according to SU(2) symmetry. Similarly, since the measurements of
branching fraction for $\jpsi\ar\nnb\pi^0$ is not currently available, 
we assume the branching fraction and its corresponding uncertainty for 
$\jpsi\ar\nnb\pi^0$ are the same as those for $\jpsi\ar\ppb \pi^0$. For the 
$\jpsi\ar\nnb$ background, the main uncertainty is from the difference
between data and MC simulation in the $\nnb$ rejection rate based on the
requirement of zero clusters in the BSC. The distribution of the
number of clusters associated with $\nnb$ is obtained from two control
samples, $\jpsi\ar\pnb +c.c.$ and $\jpsi\ar\pppr$.  First, the
distribution of the number of clusters expected for $\ppb \pi^+\pi^-
\nnb$ can be obtained by combining the $p\bar{n} \pi^-$ and $\bar{p} n
\pi^+$ samples.  Second, using the control sample, $\jpsi\ar\pppr$,
the cluster distribution for $\nnb$ can be obtained by subtracting the
$\ppb \pi^+\pi^-$ distribution from the $\ppb \pi^+\pi^- \nnb$
distribution.  The distributions of the number of clusters between
data from the control sample and MC simulation are compared, and the 
difference between MC and data for the number of $\nnb$ events with 
zero BSC clusters is a 7.6\% effect, which is taken as a systematic error. 
The estimated contributions from the peaking backgrounds are
summarized in Table~\ref{totbkg}.  The error on the expected number of
events for each peaking background includes the contributions from
both the branching fraction uncertainty and the estimated systematic
error.  For $J/\psi \rightarrow \LL$ and $\ppb$, the dominant
uncertainty arises from the geometrical acceptance and their 
associated errors are entirely correlated.

\btbl 
\caption{Expected number of events ($N_{bg}$) and 
efficiencies for peaking backgrounds.}
\bcl
\doublerulesep 2pt
\begin{tabular}{l|c|c}\hline\hline
Background channel&Efficiency&Expected\\
($\psipto,\jpsi\ar$)&(\%)&$N_{bg}$\\\hline
$\MM$&0.964&$2543\pm 254$\\
$\EE$&0.907&$2393\pm 240$\\
$\nnb$&10.46&$1011\pm 85$\\
$\ppb$&0.434&$42\pm 13$\\
$\nnb\pi^0$&0.486&$29\pm 10$\\\hline
Total&&$6018\pm 514$\\\hline
\end{tabular}
\label{totbkg}
\ecl
\etbl

A $\chi^2$ fit is used to extract the number of $\jpsi$ events in the
distribution of mass recoiling against the $\pp$ in the range $3.0$
GeV/$c^2<M^{\mbox{recoil}}_{\pp}< 3.2$ GeV/$c^2$. Here, the shape used to
describe the signal comes from the $\pp$ recoil mass spectrum from
the $\psipto,\jpsi\ar\MM$ control sample, and not from the MC
simulation due to differences between the data and MC simulation (see
Fig.~\ref{ppuu}).  The Class I background is represented by a
second-order polynomial.  The fit, with $\chi^2/ndf=62.6/75$ and shown
in Fig.~\ref{fit}, yields $6424\pm 137$ events in the peak, which
includes the contributions from both signal and peaking
backgrounds, since they have the same probability density functions
(PDF) in the fit. After subtracting the expected backgrounds listed in
Table~\ref{totbkg} from the fitted yields, we get the number of
$\psipto,\jpsi\ar \mbox{invisible} $ events to be $406\pm 532$.

The estimated uncertainties that do not cancel in the ratio are 
described here.
The systematic uncertainty caused by the tracking
efficiency in the MDC for the two muons in the control sample is
estimated to be 4\%~\cite{simbes}. The systematic error for the
requirement on $E_{tot}$ is 1.7\%, and the uncertainty
caused by the $R$ requirement in the selection of the control sample,
$\psipto,\jpsi\ar\MM$, is determined to be 1.0\% by a comparison with
and without the $R$ requirement. The uncertainty from the
background shape, which is used in the fit to
$M_{\pp}^{\mbox{recoil}}$ for Class I, is found to be negligible.
The uncertainty associated with bin-size and the range of the fit 
is 1.3\%.
In order to investigate the uncertainty caused by the trigger, we
check the four trigger channels used in the BES-II experiment. 
The largest systematic error comes from uncertainties in modeling the energy
threshold requirement in the BSC.  A conservative estimate, based on MC 
simulation, is that this corresponds to at most a 1\% uncertainty in the 
detection efficiency. The uncertainty of ${\cal B}(\jpsi\ar\MM)$ is
taken from the PDG~\cite{pdg}. The total uncertainty, which is determined by
the sum of all sources in quadrature, is 4.9\%. Taking this systematic
uncertainty into account, the upper limit for the number of events of
$\psipto,\jpsi\ar \mbox{invisible}$ is 1285 at the 90\% confidence level,
or a central value of $406^{+539}_{-333}$ at the 68.3\% confidence level
from the Feldman-Cousins frequentist approach~\cite{fc}.

Finally, the upper limit on the ratio of ${\cal B}(\jpsi\ar$
\mbox{invisible}) to ${\cal B}(\jpsi\ar\MM)$ is determined from the
relation
\beqr
\frac{{\cal B}(\jpsi\ar
\mbox{invisible})}{{\cal B}(\jpsi\ar\MM)}
&<&\frac{N^{\jpsi}_{UL}/\epsilon_{\mbox{invisible}}}
{N_{\MM}^{\jpsi}/\epsilon_{\MM}^{\jpsi}}\nonumber\\
&=&1.2\times 10^{-2},
\eeqr
where $N^{\jpsi}_{UL}$(1285) is the 90\% confidence-level upper
limit on the number of
$\psipto,\jpsi\ar \mbox{invisible}$ events,
$\epsilon_{\mbox{invisible}}(36.8\%)$ is the MC-determined signal
efficiency, $N_{\MM}^{\jpsi}(43429\pm 208)$ is the number of events
for $\psipto,\jpsi\ar\MM$ and $\epsilon_{\MM}^{\jpsi}(14.5\%)$ is the
detection efficiency for that decay mode. In addition, the two-sided
interval of the number of the measured events for invisible decays, 
$N^{\jpsi}_{invisible}$, is (73, 945) at the 68.3\%
confidence level. The corresponding two-sided interval of the ratio 
$\frac{{\cal B}(\jpsi\ar \mbox{invisible})}{{\cal B}(\jpsi\ar\MM)}$ is 
$(0.66\times 10^{-3}, ~8.6\times 10^{-3})$.

In summary, we performed the first search for invisible decays of 
the $\jpsi$ using $\psipto$ events detected in a sample of 
14.0 million $\psp$ decays.
The upper limit on the ratio $\frac{{\cal B}(\jpsi\ar
\mbox{invisible})}{{\cal B}(\jpsi\ar\MM)}$ at the 90\% confidence level is 
$1.2\times 10^{-2}$.  This measurement improves by a factor of 3.5 the 
bound on the product of the coupling of the $U$ boson to the $c$ quark and 
LDM particles as described in Eqs. (25) and (26) of Ref.~\cite{00}. One now 
has, for a Majorana LDM particle $\chi$ as in Eq. (26) of Ref.~\cite{00}, a 
limit of $|c_{\chi}f_{cV}|< 8.5\times 10^{-3}$, which is almost a factor of 
2 stronger than the corresponding limit $|c_{\chi}f_{bV}|<1.4\times
10^{-2}$ derived from the invisible decays of the $\Upsilon(1S)$
as described in Eq. (106) in Ref.~\cite{000}, where $c_{\chi}$ and
$f_{cV}$ ($f_{bV}$) denote the $U$ boson couplings to the LDM
particle $\chi$ and $c$ ($b$) quark. We expect a more precise
measurement can be obtained in the future BES-III experiment.

The BES collaboration thanks the staff of BEPC and computing
center for their hard efforts. We would also thank P.~Fayet,
R.~McElrath and S.~H.~Zhu for useful discussions and suggestions.
This work is supported in part by the National Natural Science Foundation of
China under contracts Nos. 10491300, 10225524, 10225525, 10425523,
the Chinese Academy of Sciences under contract No. KJ 95T-03, the
100 Talents Program of CAS under Contract Nos. U-11, U-24, U-25,
and the Knowledge Innovation Project of CAS under Contract Nos.
U-602, U-34 (IHEP), the National Natural Science Foundation of
China under Contract No. 10225522 (Tsinghua University), the
Swedish research Council (VR), and the Department of Energy under
Contract No.DE-FG02-04ER41291 (U Hawaii).

\end{document}